\documentclass[dvips]{acta}
\usepackage{supertabular,lscape,epsfig}
\usepackage{amssymb}
\usepackage{amsmath}
\DeclareSymbolFont{ppa}{OT1}{ppl}{m}{it}
\DeclareMathSymbol{\vv}{\mathalpha}{ppa}{'166}

\newfont{\hb}{rphvb at 10pt}
\newfont{\hbo}{rphvbo at 10pt}
\newfont{\bitt}{rptmbi at 12pt}
\newfont{\bits}{rptmbi at 11pt}

\SetPages{1}{1}

\SetVol{66}{2016}

\begin{document}

\newcommand{\TabCapp}[2]{\begin{center}\parbox[t]{#1}{\centerline{
  \small {\spaceskip 2pt plus 1pt minus 1pt T a b l e}
  \refstepcounter{table}\thetable}
  \vskip2mm
  \centerline{\footnotesize #2}}
  \vskip3mm
\end{center}}

\newcommand{\TTabCap}[3]{\begin{center}\parbox[t]{#1}{\centerline{
  \small {\spaceskip 2pt plus 1pt minus 1pt T a b l e}
  \refstepcounter{table}\thetable}
  \vskip2mm
  \centerline{\footnotesize #2}
  \centerline{\footnotesize #3}}
  \vskip1mm
\end{center}}

\newcommand{\MakeTableSepp}[4]{\begin{table}[p]\TabCapp{#2}{#3}
  \begin{center} \TableFont \begin{tabular}{#1} #4
  \end{tabular}\end{center}\end{table}}

\newcommand{\MakeTableee}[4]{\begin{table}[htb]\TabCapp{#2}{#3}
  \begin{center} \TableFont \begin{tabular}{#1} #4
  \end{tabular}\end{center}\end{table}}

\newcommand{\MakeTablee}[5]{\begin{table}[htb]\TTabCap{#2}{#3}{#4}
  \begin{center} \TableFont \begin{tabular}{#1} #5
  \end{tabular}\end{center}\end{table}}

\newfont{\bb}{ptmbi8t at 12pt}
\newfont{\bbb}{cmbxti10}
\newfont{\bbbb}{cmbxti10 at 9pt}
\newcommand{\uprule}{\rule{0pt}{2.5ex}}
\newcommand{\douprule}{\rule[-2ex]{0pt}{4.5ex}}
\newcommand{\dorule}{\rule[-2ex]{0pt}{2ex}}
\def\thefootnote{\fnsymbol{footnote}}
\begin{Titlepage}
\Title{Gaia and Variable Stars}
\Author{A.~~U~d~a~l~s~k~i$^1$,~~
I.~~S~o~s~z~y~ñ~s~k~i$^1$,~~
D.\,M.~~S~k~o~w~r~o~n$^1$,~~
J.~~S~k~o~w~r~o~n$^1$,\\
P.~~P~i~e~t~r~u~k~o~w~i~c~z$^1$,~~
P.~~M~r~ó~z$^1$~~
R.~~P~o~l~e~s~k~i$^{1,2}$,~~\\
M.\,K.~~S~z~y~m~a~ñ~s~k~i$^1$,~~
S.~~K~o~z~³~o~w~s~k~i$^1$,~~
\L.~~W~y~r~z~y~k~o~w~s~k~i$^1$,\\
K.~~U~l~a~c~z~y~k$^{1,3}$~~
and~~M.~~P~a~w~l~a~k$^1$}
{$^1$Warsaw University Observatory, Al.~Ujazdowskie~4, 00-478~Warszawa, Poland\\
e-mail: (udalski,soszynsk)@astrouw.edu.pl\\
$^2$ Department of Astronomy, Ohio State University, 140 W. 18th Ave., Columbus, OH 43210,~USA\\
$^3$ Department of Physics, University of Warwick, Gibbet Hill Road, Coventry, CV4 7AL,~UK}
\end{Titlepage}

\Abstract{

We present a comparison of the Gaia DR1 samples of pulsating variable
stars -- Cepheids and RR Lyrae type -- with the OGLE Collection of
Variable Stars aiming at the characterization of the Gaia mission
performance in the stellar variability domain.

Out of 575 Cepheids and 2322 RR Lyrae candidates from the Gaia DR1
samples located in the OGLE footprint in the sky, 559 Cepheids and 2302
RR Lyrae stars are genuine pulsators of these types. The number of
misclassified stars is low indicating reliable performance of the Gaia
data pipeline.

The completeness of the Gaia DR1 samples of Cepheids and RR Lyrae stars
is at the level of 60--75\% as compared to the OGLE Collection dataset.
This level of completeness is moderate and may limit the applicability
of the Gaia data in many projects.} {Stars: variables: Cepheids --
variables: RR Lyrae -- Magellanic Clouds}

\Section{Introduction}

Gaia is the prime astrophysical space mission of the European Space
Agency. Its main scientific objective is to provide precise 3-D maps of
the Galaxy. To achieve this goal, the satellite is continuously scanning
the entire sky for a period of at least five years with its astrometric,
photometric and spectroscopic instruments. 

Precise astrometry, \ie stellar parallaxes and stellar proper motions,
of about billion Milky Way stars, is the most awaited science outcome of
the Gaia mission. However, the mission was also designed to provide
precise wide band photometry of stars down to about 20th magnitude and
low resolution spectral information as well as high accuracy radial
velocities for brighter objects. Details of the mission can be found in
Gaia Collaboration \etal (2016a).

Although the Gaia mission has mostly been focused on astrometry, because
this is the main field of astrophysics still lacking the
state-of-the-art modern observations, the expectation for the Gaia
photometric and spectroscopic part of the program have also been high.
For example, Eyer \etal (2012) predicted the detection and
characterization of millions of new variable stars in the course of the
mission. 

After the successful launch in December 2013, the Gaia mission entered
its commissioning phase. One of the elements of this test phase was a
special sky scanning mode (Ecliptic Pole Scanning Law, EPSL) lasting for
about one month (28 days) and enabling frequent monitoring of the sky
regions around the North and South Ecliptic Poles.

The South Ecliptic Pole is located on the outskirts of the Large
Magellanic Cloud. This is the region of the sky has regularly been
monitored since 2010 by the Optical Gravitational Lensing Experiment
(OGLE) -- a long-term sky-variability survey (Udalski, Szyma{\'n}ski and
Szyma{\'n}ski, 2015). Being aware of the Gaia plans for the
commissioning phase tests, the OGLE team prepared a set of ground based
benchmarks for testing the Gaia satellite performance in practice
(Soszy{\'n}ski \etal 2012). Four OGLE fields around the Gaia South
Eclipsing Pole (GSEP) were extensively investigated and photometric
maps, variable stars, galaxies, high proper motion stars and generally
proper motions for Galactic foreground stars in these fields were
presented.

On September 14, 2016 the Gaia team has released the first observational
data set from the mission -- DR1 (Gaia Collaboration \etal 2016b). The
Gaia DR1 primarily contains astrometry for brighter stars (based on
Hipparcos satellite measurements obtained in 1990s and Gaia present
ones) as well as positions and magnitudes of Gaia detected objects.
Additionally, the results of an analysis on the pulsating variable stars
based mainly on the commissioning phase EPSL data from the South
Ecliptic Pole were also released as a part of the Gaia DR1 (Gaia
Collaboration \etal 2016c).

Gaia DR1 enables a direct comparison of the mission performance and
efficiency to the ground based OGLE data. It is not yet possible to test
astrometric results at this stage because the stars measured by OGLE in
GSEP are much fainter than the stars with full astrometry released in
DR1. However, the release of stellar variability results makes it
possible to carefully analyze the mission performance and draw more
reliable conclusions on the Gaia's final outcome in this field. 

Here we present a comparison of the Gaia variable stars from DR1 with
the most complete existing ground based dataset of variable stars -- the
OGLE Collection of Variable Stars (Soszy{\'n}ski \etal 2014, 2015ab,
2016). The latter dataset has over 90\% completeness, thus can be used
as an almost perfect reference point of the analysis. Long-term OGLE
light curves and a high photometric precision allow reaching high
classification purity with practically no false positives which happen
in poorly sampled data.
 
One has to remember, however, that the Gaia commissioning EPSL data are
much better suited for a variability search than the data collected during
the regular Gaia observations (Nominal Scanning Law, NSL) due to much
higher observing cadence. Thus, the results presented here will rather
be the upper limits of what can be expected from the mission.

\Section{Gaia and OGLE Observational Data}

The Gaia data on variable stars contain observations mostly from the
commissioning EPSL phase supplemented with several epochs collected
during the first months of regular operation. They were included in the
Gaia DR1 (Gaia Collaboration \etal 2016b) and can be downloaded from
{\it http://archives.esac.esa.int/gaia/}. The dataset includes
photometry in the Gaia photometric G-band, position in the sky, periods,
pulsation characteristics and classification from the data pipeline. The
detailed description of the data pipeline can be found in Gaia
Collaboration \etal (2016c). Released data contain two basic types of
pulsating variable stars -- Cepheids and RR Lyrae stars. The former set
includes anomalous, Type-II and classical Cepheids. The final result of
the analysis consists of 599 and 2595 candidates for Cepheids and RR
Lyrae stars, respectively,  detected in this Gaia dataset. 43 Cepheids
and 343 RR Lyrae stars are claimed to be new discoveries. 

The OGLE Collection of Variable Stars (OCVS) is based on the data
collected during the fourth phase of the OGLE survey, OGLE-IV (Udalski
\etal 2015) and supplemented by earlier discoveries from the previous
phases of the OGLE project: OGLE-III Catalog of Variable Stars
(Soszy{\'n}ski \etal 2013 and references therein), OGLE-II, and OGLE-I
catalogs. For the selected fields precise OGLE photometry spans almost
25 years. In total, the OCVS now contains almost one million classified
variable stars.

The OCVS sections on pulsating stars in the Magellanic System (the
Magellanic Clouds and Magellanic Bridge) have already been released in a
series of three papers on anomalous Cepheids (Soszy{\'n}ski \etal
2015a), classical Cepheids (Soszy{\'n}ski \etal 2015b), and RR Lyrae
stars (Soszy{\'n}ski \etal 2016). The OCVS section on Population II
Cepheids (T2CEPs) has not been published yet, however it is in an
advanced stage of preparation and will be released in 2017.

The OGLE-IV 32-CCD mosaic camera contains technical gaps between
detectors (Udalski \etal 2015). This dead area can still be filled by
observations thanks to natural dithering caused by imperfections of the
telescope pointing and small artificial dithering introduced during
selected observing seasons. However, only a small part of dead area is
filled on the standard OGLE-IV reference images, on which the current
OCVS photometry is based. Thus, about 7\% of the each field covered by
the OGLE-IV pointing falls into this ``dead zone'', lowering the
completeness.

This is not a problem in the regions that were observed during the
previous phases of the OGLE survey. Older detections nicely fill the
OGLE-IV gaps. These are, for example, central parts of the Large and
Small Magellanic Clouds where the completeness of Cepheid detection is
close to 100\% and in the case of fainter RR Lyrae stars well over 90\%.

As the existing ``dead zones'' limit the OGLE completeness and
interesting objects can be missed in spite of having observing data, a
special set of new deep reference images has been constructed for
reaching much deeper magnitudes and masking practically all ``dead
zones''. They are composed of 50--100 individual good resolution
individual images and reach almost three magnitudes fainter stars in
uncrowded fields.

Additional photometric reductions with the new deep reference image sets
of the Magellanic System fields were carried out to extract variable
objects located on the ``dead zones'' in the fields. Final results of
this search will be the base of the OCVS update and extension which is
planned for 2017.

\Section{Gaia Variable Stars Classification}

Regularly pulsating variable stars such as Cepheids or RR Lyrae are
relatively easy to detect, especially those pulsating in the fundamental
mode (RRab for RR Lyrae, F for Cepheids). Their light curves have
characteristic shape and relatively large amplitudes in the optical
bands. The situation is much different in infra-red bands -- the light
curves of these pulsating stars become more sinusoidal and of lower
amplitude and can be easily misclassified. On the other hand, stars
pulsating in overtone modes (RRc, 1O or 2O Cepheids) also often have
quasi-sinusoidal light curves even in optical bands. Therefore they
should be classified with even more caution.

To assess the Gaia dataset of Cepheids and RR Lyrae stars presented in
Gaia DR1 (Gaia Collaboration \etal 2016c) we cross-identified the sample of 3194
variable stars presented on the final Gaia pipeline list (599 Cepheid
and 2595 RR Lyrae candidates) with the OGLE detected objects using
RA/DEC coordinates provided within Gaia DR1. First, we checked which of
the Gaia candidates fall into the OGLE-IV field footprint in the sky. It
turned out that OGLE should see 575 objects out of 599 Gaia Cepheid
candidates (96\%). Similar statistics for RR Lyrae stars are 2322
objects out of 2595 Gaia candidates (89.5\%). 

The field covered by Gaia is located in the northern part of the LMC and
this region of the sky is covered by OGLE practically only during the
OGLE-IV phase. Thus, as we already mentioned, the completeness of the
currently released OCVS in this region suffers from ``dead zones''
between the mosaic camera CCDs. To minimize this bias in further
analysis we also used unpublished data from planned extension of the
OCVS covering these gaps and also unpublished yet OGLE detections of
T2CEPs.

In the next step, we performed object by object cross-identification. We
conservatively used the search radius of 2\arcs, as the coordinate
shifts for a few stars exceeded 1\arcs.

517 objects out of 575 Gaia Cepheid candidates lying in the OGLE-IV
fields were found in the current release of the OCVS. This number
includes three OGLE RR Lyrae stars which Gaia pipeline classified as
Cepheid candidates. After careful checking of the OGLE unpublished data,
the total number of positive Cepheid cross-identifications increased to
556. Only three stars out of 575 could not be assessed. Two are missing
in all OGLE databases because they are located very close to bright
overexposed stars and one has only nine observations -- too few for a
verification. Their Gaia light curves suggest, however, that all of them
are indeed Cepheids. Thus, the total number of confirmed Cepheids detected
by Gaia in the OGLE-IV footprint is 559. 16 Gaia Cepheid candidates have
been misclassified. Table~1 lists these objects. Their Gaia and OGLE-IV
light curves are shown in Appendix~A.

\MakeTableee{crcc}{12.5cm}{Misclassified variables from the Gaia lists
located in the OGLE-IV footprint in the sky}
{\hline
\noalign{\vskip3pt}
Gaia ID & \multicolumn{1}{c}{Period} & G     & OGLE type \\
        & \multicolumn{1}{c}{[days]} & [mag] &           \\
\noalign{\vskip3pt}
\hline
\multicolumn{4}{c}{Cepheids}\\
\hline
\noalign{\vskip3pt}

4658925092406745984 & 2.69331 & 17.372	& ELL/ECL\\
4658950381175117824 & 2.79569 & 15.151	& ELL/ECL\\
4659497285129779584 & 12.3449 & 17.271	& ELL/ECL\\
4659525872450052480 & 16.2036 & 16.877	& ELL/ECL\\
4660248221506229248 & 7.73603 & 17.360	& ELL/ECL\\
4661973149098711296 & 1.60573 & 18.841	& ELL/ECL\\
4661995066309511296 & 18.4069 & 16.926	& ELL/ECL\\
4662006679902148224 & 7.70166 & 16.806	& ELL/ECL\\
4662348181341882368 & 11.4239 & 17.694	& ELL/ECL\\
4663540498623553152 & 3.77478 & 17.719	& ELL/ECL\\
4663548916758167168 & 3.00257 & 17.260	& ELL/ECL\\
4662029666568214400 & 0.78565 & 17.080	& SPOTTED\\
4662232732622395008 & 3.03502 & 17.323	& SPOTTED\\
4663192228315671808 & 3.09877 & 16.008	& SPOTTED\\
4663913885900050560 & 0.84573 & 17.522	& SPOTTED\\
5283282976398012160 & 32.1046 & 15.479	& SRV\\
\noalign{\vskip3pt}
\hline
\multicolumn{4}{c}{RR Lyrae stars}\\
\hline
\noalign{\vskip3pt}
4659944859379434240 & 0.27130 & 19.317	& ELL/ECL\\
4660214342840128000 & 0.25803 & 19.351	& ELL/ECL\\
4660424757548266368 & 0.34667 & 18.534	& ELL/ECL\\
4660477744584252672 & 0.26666 & 19.517	& ELL/ECL\\
4660540897781466240 & 0.39683 & 19.868	& ELL/ECL\\
4660684933800980224 & 0.36905 & 18.829	& ELL/ECL\\
4660709844592511744 & 0.26240 & 19.364	& ELL/ECL\\
4660958226847368832 & 0.37466 & 18.953	& ELL/ECL\\
4660968607781569536 & 0.35880 & 19.091	& ELL/ECL\\
4662575848968098048 & 0.41629 & 19.815	& ELL/ECL\\
4662968477701040640 & 0.43160 & 19.056	& ELL/ECL\\
4663383440258523520 & 0.28563 & 19.373	& ELL/ECL\\
4663493391439108608 & 0.42364 & 19.266	& ELL/ECL\\
4663698798239074688 & 0.34272 & 19.895	& ELL/ECL\\
4664646882203956992 & 0.28848 & 19.152	& ELL/ECL\\
5284186946756087936 & 0.24678 & 18.611	& ELL/ECL\\
4675190889470416768 & 0.40385 & 18.994	& ELL/ECL\\
4659691280246136704 & 0.36279 & 18.971	& ELL/ECL\\
4675441200165453312 & 0.23730 & 19.205	& ELL/ECL\\
4663719203629897856 & 0.72300 & 19.032	& ELL/ECL\\
\noalign{\vskip3pt}
\hline}

Similar figures for Gaia RR Lyrae candidates are as follows: 2143
objects out of 2322 Gaia candidates located in the OGLE-IV footprint
were found in the current release of the OCVS. After checking OGLE-IV
unpublished data, the number of positive cross-identifications increased
to 2283. Only 19 stars could not be verified in the OGLE databases. 13
of them had too small number of measurements ($N<15$) to assess the
light curve. The remaining six include two very bright Galactic RR Lyrae
that are overexposed on OGLE images and the four remaining are located
in the still present, very tiny ``dead zones''. Gaia light curves
indicate that all of these 19 objects are true RR Lyrae stars. Thus, the
total number of the Gaia genuine RR Lyrae stars in the OGLE-IV footprint
is 2302 out of 2322 candidates. The missclasified objects from the Gaia
RR Lyrae star list are also listed in Table~1. Appendix~B presents their
Gaia and OGLE-IV light curves.

Our comparison indicates that the classification of the Gaia pulsating
stars dataset is generally correct. This is not surprising -- Cepheids
and RR Lyrae stars in the optical band are easy to detect, as already
mentioned. Additionally, the Gaia EPSL phase sampling was much more
favorable for the detection of these short period variables than that
used during the main mission. 

It is more interesting to check the classification failures. These are,
for example, quasi-sinusoidal light curves of ellipsoidal stars or
eclipsing binaries which can be properly assessed only when having large
number of observations what reveals subtle effects (\eg non-equal depth
minima). Several objects show changing pattern of variability at
long-term scales -- likely due to spots -- not pulsations. In a single
case the candidate is a semi-regular variable -- not revealing periodic
variability noted by Gaia.

\Section{Completeness of the Gaia Variable Stars}

The variable stars are one of the basic tools for studying the structure
of the environment they live in. It is crucial, then, to know the
completeness of their samples before undertaking any deeper analysis.
The OGLE-IV very high completeness ($>90$\%) samples of pulsating stars
in the Magellanic System are ideal for verification of the Gaia dataset.

The part of the sky covered by Gaia EPSL phase observations has a
specific shape and crosses several OGLE-IV fields. It is not trivial to
set precise boundaries of the Gaia scanning area in the OGLE fields.
Thus, to perform the complenteness test we decided to use only the
fields which sit fully in the Gaia covered region: LMC519, LMC506
(except corner subfields '26' and '27'), LMC512, LMC534, and LMC541
(Fig.~1). There are a few more OGLE fields fully filled by the Gaia
scanning area but they are in the outskirts of the LMC where the number
of pulsating stars is small and the results of the comparison would be
non-representative.

\begin{figure}[htb]
\begin{center}
\includegraphics[width=12.5cm,bb=-158 -166 770 960 ]{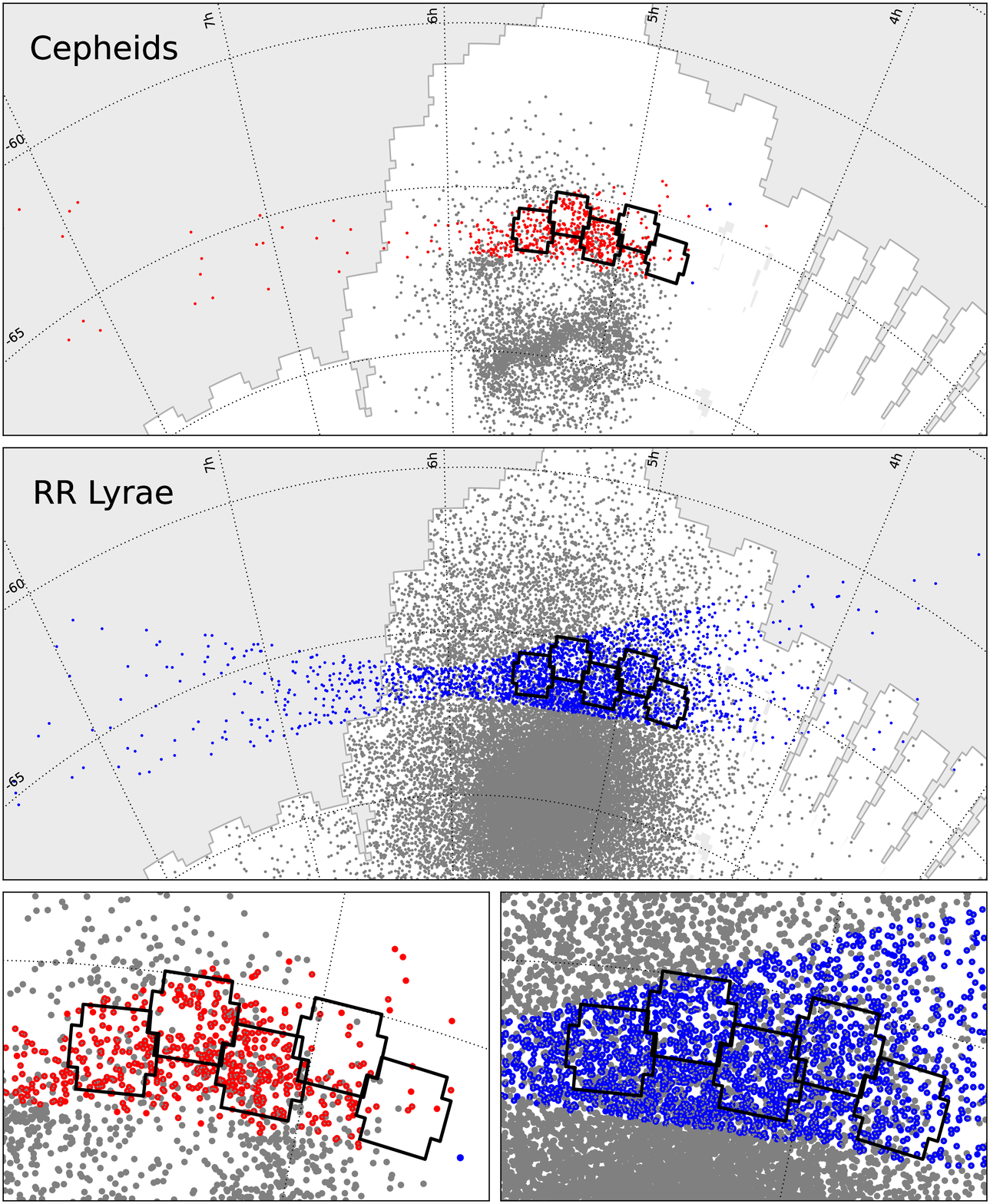}
\end{center}
\FigCap{{\it Upper panel:} sky coverage of the Gaia DR1 sample of
Cepheids (red dots). Gray dots mark positions of the OGLE-IV Cepheids.
{\it Midle panel:} sky coverage of the Gaia DR1 sample of  RR Lyrae
stars (blue dots). Gray dots mark positions of the OGLE-IV RR Lyrae
stars. {\it Bottom panel:} close-up of the region where the completeness
of the Gaia samples was analyzed. OGLE-IV fields outlined with thick
lines are: LMC519, LMC506, LMC512, LMC534, and LMC541 (from the left to
right). Sky region outside the OGLE-IV footprint in the sky is gray
shaded.}
\end{figure}

We extracted Cepheids and RR Lyrae stars located in each of these
OGLE-IV fields from the OCVS. We supplemented them with the detections
from the planned OCVS extension to minimize OGLE-IV camera gaps
incompleteness. Finally, we extracted Cepheids/RR Lyrae stars located in
the selected OGLE-IV fields from the list of the Gaia Cepheids/RR Lyrae
positively cross-identified with the OGLE genuine pulsators.

\MakeTableee{ccccc}{12.5cm}{Completeness of the Gaia DR1 samples of
Cepheids and RR Lyrae stars}
{\hline
\noalign{\vskip3pt}
OGLE-IV Field & Average number & Number of verified & Number in &  Gaia completenss \\
              & of Gaia epochs & Gaia pulsators     & OCVS      &     [\%]\\
\noalign{\vskip3pt}
\hline
\multicolumn{5}{c}{Cepheids}\\
\hline
\noalign{\vskip3pt}
LMC519	&	103	&	80	&	106	&	75.5\\
LMC506  &	75	&	97	& 	127	& 	76.4\\
LMC512	&	66	&	119	&	162	&	73.5\\
LMC534	&	58	&	19	&	31	&	61.3\\
LMC541	&	53	&	6	&	11	&	54.5\\
\noalign{\vskip3pt}
\hline
\multicolumn{5}{c}{RR Lyrae stars}\\
\hline
\noalign{\vskip3pt}
LMC519	&	98	&	225	&	333	&	67.6\\
LMC506  &	73	&	164	& 	218	& 	75.2\\
LMC512	&	67	&	218	&	325	&	67.1\\
LMC534	&	58	&	136	&	217	&	62.7\\
LMC541	&	50	&	85	&	136	&	62.5\\
\noalign{\vskip3pt}
\hline}

Table~2 presents the results of our study of completeness. As can be
seen the completeness of the Gaia DR1 sample is moderate at the level of
60--75\% for both -- Cepheids and RR Lyrae stars. Additionally, in
Table~2 we also list the average number of epochs collected by Gaia for
both types of pulsators in each of the fields.

The completeness of detection for both types of pulsating stars is
generally similar and does not seem to be very strongly dependent on the
number of collected epochs. Generally, Cepheid variables are
significantly brighter than RR Lyrae stars, thus differences in
completeness of these two types of stars may provide information on the
Gaia performance depending on the brightness of the object.

\Section{Discussion}

We have compared the Gaia samples of pulsating star candidates, Cepheids
and RR Lyrae stars, released with the Gaia DR1 (Gaia Collaboration \etal
2016c) with the OGLE Collection of Variable Stars. The Gaia data come
mostly from the commissioning EPSL phase. The main conclusion from this
comparison is that while the classification of pulsating stars is sound
in this data sample, the completeness of the sample is moderate at the
level of only 60--75\% level for Cepheids and RR Lyrae stars. This may
limit the applicability of the Gaia data for more complex analyses, for
example of the Magellanic Clouds structure.

It is important to remember that the Gaia DR1 sample is not fully
representative of the Gaia variable stars outcome from the main mission.
Much lower cadence during regular NSL observations -- on average only
about 70 epochs per 5 years (except for limited regions in the sky) and
in practice much less as some of them will be very close in time -- will
make the detection and proper characterization of many variable stars
much more difficult than during the EPSL phase. Thus, the numbers from
our analysis should be treated rather as upper limits of what Gaia can
achieve in the variable stars domain of the mission.

For example, one of the main problems with the proper classification of
pulsating stars can be variables that mimic light curves of pulsating
stars. The LMC instability strip where Cepheids and RR Lyrae stars are
located, thus the Gaia DR1 sample, is only moderately contaminated by
the Galactic foreground stars of late spectral type. In the other sky
regions, in particular in the Galactic disk the contamination will be
much more severe.  

Using solely Fourier parameters of light curves for classification can
often be misleading in such cases as the parameters of contaminators can
be in the typical pulsating stars range. Additional tools for the
classification are of limited use for the Milky Way objects. For
example, the color and magnitude information affected by unknown amount
of reddening and unknown distance will not provide such important
constraints on the color-magnitude object location as in the case of the
Magellanic Clouds or star clusters.

\begin{figure}[htb]
\begin{center}
\includegraphics[width=12.5cm,bb=70 50 540 640]{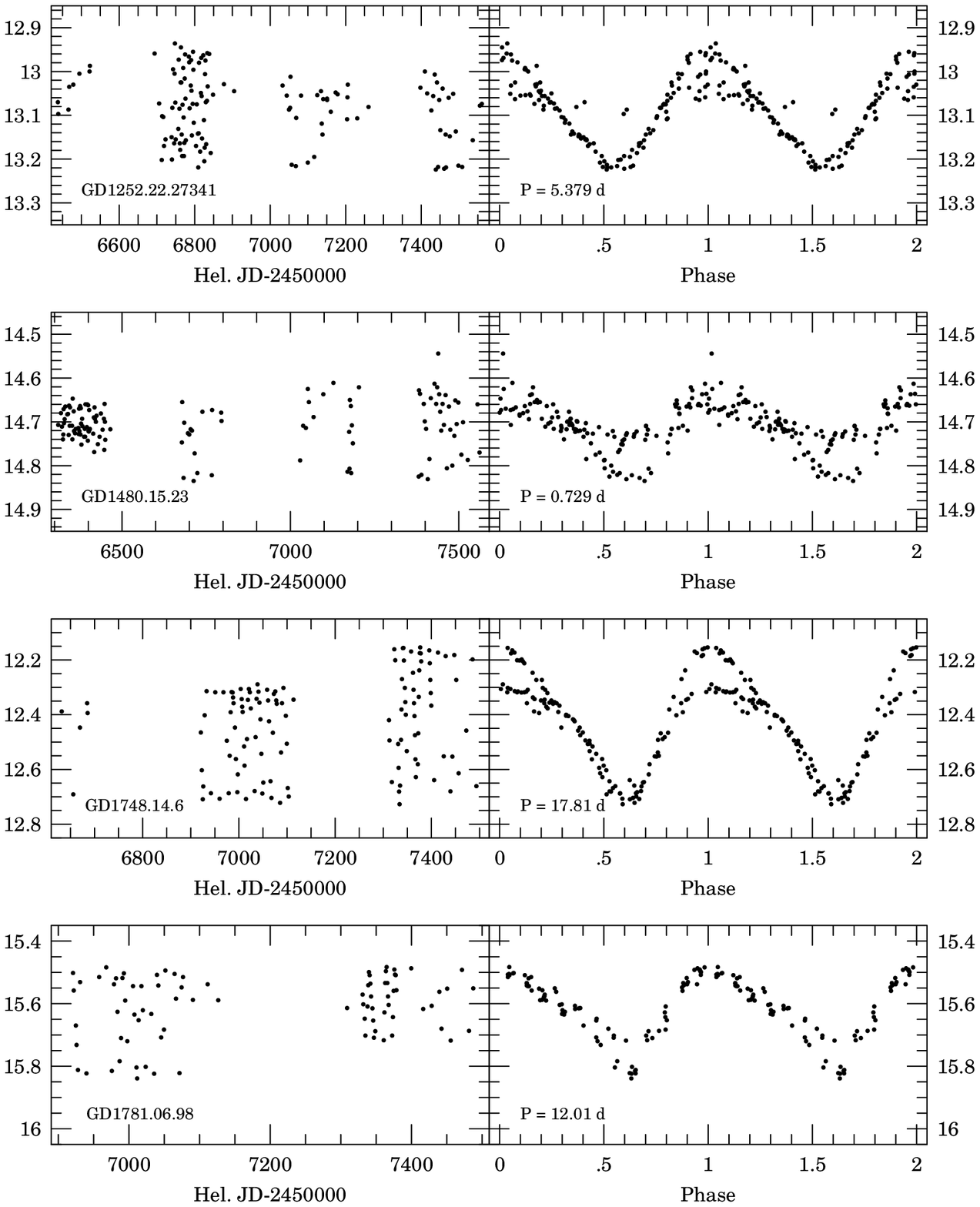}
\end{center}
\FigCap{Examples of spotted stars mimicking pulsating variable stars 
from the OGLE-IV Galaxy Variability Survey.}
\end{figure}

A sample of such stars mimicking Cepheid variables from the OGLE
Galactic plane survey is shown in Fig.~2. These are in fact spotted
stars. Their evolution of light curves can be easily seen thanks to
long-term well-sampled observations. In extreme cases variability ceases
on a month time scale. Another examples of variables mimicking Cepheids
can be found in Pietrukowicz \etal (2015).  Such objects are very
numerous in the Galaxy and certainly may contaminate the Gaia outcome to
large extent.

24 Cepheid and 273 RR Lyrae candidates selected from Gaia  EPSL phase
data are located outside the OGLE-IV footprint. While the distribution
of RR Lyrae stars in the sky looks as expected (red dots in Fig.~32 of
Gaia Collaboration \etal 2016c) -- they are the LMC halo objects -- the
distribution of Cepheid candidates looks very suspicious (Fig.~35
there). They form a long strip along the scanning path where they are
relatively uniformly distributed (\cf Fig.~1). 

This is highly surprising because the OGLE-IV fields include a similar
strip -- five degree wide in declination and extending even farther to
the East but about five degrees South in declination. No Cepheids were
found in this large area. This prompted us to take a closer look at the
Gaia Cepheid and RR Lyrae candidates from this region.

We carefully inspected the original Gaia light curves of 24 Cepheid
candidates and applied the same classification criteria we use for the
OGLE Collection. Only four stars from this sample survived this test,
IDs: 5280412430710766080, 5281522211604337920, 5282067225775482880,
5283779131019074048. Objects 5288333583059787264 and 5281307222723913472
may also be Cepheids but a much longer dataset is needed to confirm
their nature (especially that both these objects are far east from the
LMC and may be simply Galactic spotted contaminants). Five
additional pulsating-like stars in this sample are RR Lyrae stars and
the remaining objects are rather contaminants than genuine Cepheids and
we list them in Table~3. 

\MakeTableSep{crcc}{12.5cm}{Misclassified variables from the Gaia lists
located outside the OGLE-IV footprint in the sky}
{\hline
\noalign{\vskip3pt}
Gaia ID & \multicolumn{1}{c}{Period} & G     & Most likely \\
        & \multicolumn{1}{c}{[days]} & [mag] &   type    \\
\noalign{\vskip3pt}
\hline
\multicolumn{4}{c}{Cepheids}\\
\hline
\noalign{\vskip3pt}
5289779853168752384 & 2.57603   & 15.497	& ELL/ECL\\
5274727126666346240 & 1.28488   & 16.902	& ELL/ECL\\
5278062426470300288 & 4.63284   & 15.445 	& ELL/ECL\\
5280952252264821760 & 1.09380   & 16.522	& ELL/ECL\\
5281186997996634624 & 6.02057   & 14.003	& ELL/ECL\\
5282252871440640896 & 12.0150  &  13.934	& ELL/ECL\\
5283498927352479488 & 0.28700   & 17.950	& ELL/ECL\\
5283638599695956608 & 33.2353  &  15.477	& ELL/ECL\\
5283813318959448320 & 0.33173   & 18.037 	& ELL/ECL\\
5285121669077034496 & 6.96617   & 15.253	& ELL/ECL\\
5288371722370099968 & 1.38654   & 17.649	& ELL/ECL\\
5289741404620290688 & 1.22146   & 18.726	& ELL/ECL\\
5289507105564319616 & 17.8006  &  15.858	&  UNDEF\\
\noalign{\vskip3pt}
\hline
\multicolumn{4}{c}{RR Lyrae stars}\\
\hline
\noalign{\vskip3pt}
4670192921927875584 & 0.37975    & 19.172	& ELL/ECL \\
4674761839418372480 & 0.38877    & 18.989	& ELL/ECL \\
4674767543134984448 & 0.38139    & 19.096	& ELL/ECL \\
4674938517192984576 & 0.35799    & 19.146	& ELL/ECL \\
4679938065283825280 & 0.35933    & 19.064	& ELL/ECL \\
5275325673307993728 & 0.33847    & 16.757	& ELL/ECL \\
5275475928444428416 & 0.37113    & 19.103	& ELL/ECL \\
5280392841364170112 & 0.34270    & 19.170	& ELL/ECL \\
5281340826546416896 & 0.34362    & 19.255	& ELL/ECL \\
5281414631263561728 & 0.34609    & 19.097	& ELL/ECL \\
5281874467642636288 & 0.35250    & 19.132	& ELL/ECL \\
5282132921594371968 & 0.35744    & 19.110	& ELL/ECL \\
5282395052037988992 & 0.27385    & 19.021	& ELL/ECL \\
5283362862789508480 & 0.26416    & 19.024	& ELL/ECL \\
5283393786555282816 & 0.33703    & 19.321	& ELL/ECL \\
5283586583340633088 & 0.27014    & 19.087	& ELL/ECL \\
5283621625979449600 & 0.35482    & 19.144	& ELL/ECL \\
5283633445728974592 & 0.42496    & 18.823	& ELL/ECL \\
5283637568897023232 & 0.37903    & 19.055	& ELL/ECL \\
5283730580708733824 & 0.33358    & 19.211	& ELL/ECL \\
5283735803389049472 & 0.35052    & 19.094	& ELL/ECL \\
5283758893133924736 & 0.32324    & 19.153	& ELL/ECL \\
5283795726772805632 & 0.33888    & 19.012	& ELL/ECL \\
5284416504166282496 & 0.27559    & 19.157	& ELL/ECL \\
5284416950842889856 & 0.38347    & 18.992	& ELL/ECL \\
5285008556818105216 & 0.65043    & 18.943	& ELL/ECL \\
5285101362472065536 & 0.37397    & 18.930	& ELL/ECL \\
\noalign{\vskip3pt}
\hline}

\setcounter{table}{2}
\MakeTableee{crcc}{12.5cm}{Concluded}
{\hline
\noalign{\vskip3pt}
Gaia ID & \multicolumn{1}{c}{Period} & G     & Most likely \\
        & \multicolumn{1}{c}{[days]} & [mag] &   type    \\
\noalign{\vskip3pt}
\hline
\noalign{\vskip3pt}
5285230108411033088 & 0.35623    & 19.196	& ELL/ECL \\
5285243886665938304 & 0.30257    & 15.543	& ELL/ECL \\
5285277559209271552 & 0.28858    & 19.184	& ELL/ECL \\
5285564050708109696 & 0.28067    & 19.235	& ELL/ECL \\
5287819595734156544 & 0.37067    & 19.186	& ELL/ECL \\
5288408659087559936 & 0.33325    & 19.504	& ELL/ECL \\
\noalign{\vskip3pt}
\hline}

These four remaining sound Cepheid candidates from the Gaia list are now
much more consistent with the picture of the OGLE Cepheid distribution.
Two of them are very close to the OGLE-IV fields LMC608 and LMC614 where
single Cepheids occur. The two remaining are much farther. However,
single Cepheids, in particular anomalous, have been detected at similar
distances from the LMC center in other directions.

We also carefully inspected Gaia light curves of 273 RR Lyrae candidates
outside the OGLE-IV footprint. 240 survived our visual inspection and
can be treated as bona-fide RR Lyrae stars. The remaining ones are
rather contaminants (see also Table~3).

In both Cepheid and RR Lyrae star samples from the LMC the main
contaminants are ellipsoidal/eclipsing systems that mimic first overtone
light curves of quasi-sinusoidal shape. A careful investigation of such
light curves folded with double period allows noticing unequal depth of
minima/maxima or a different shape of eclipses that immediately excludes
classification as pulsating stars. Better sampling and longer time-span
of the dataset are needed to be able to see such details and clean the
final sample from the contaminants. For example, OGLE typically starts
variable star searches after three years of observations and when the
number of epochs exceeds 100.

\Section{Summary}

Our tests of the Gaia DR1 on pulsating variable stars -- Cepheids and RR
Lyrae -- indicate that the classification of individual stars of these
types is generally done correctly. The number of misclassifications is
small. However, the LMC fields are relatively pure from possible
contaminating objects like, for example, spotted stars which are much
more common in the Galactic fields. After OGLE verification the number
of bona-fide Cepheids and RR Lyrae stars is 559 and 2302, respectively,
in the OGLE footprint. Outside the OGLE covered area 245 RR Lyrae stars
(240 from the Gaia RR Lyrae list and five from the Cepheid list) and
only four Cepheids seem to be genuine variables of these types.

On the other hand, the completeness of the sample is rather moderate --
about 60--75\% for Cepheids and RR Lyrae stars. This level of the
completeness of the Gaia variable stars data may be not sufficient in
many projects. One should also remember that the Gaia DR1 data come
mainly from the commissioning EPSL phase which was much better suited
for short period variable stars search than the standard Gaia operation.
Thus, the results of our analysis should be treated as upper limits. 

\Acknow{
We would like to thank Profs. M. Kubiak and G. Pietrzy{\'n}ski, former
members of the OGLE team, for their contribution to the collection of
the OGLE photometric data over the past years.

The OGLE project has received funding from the Polish National Science
Centre grant MAESTRO no. 2014/14/A/ST9/00121 to AU. This work has been
supported by the Polish Ministry of Science and Higher Education through
the program ``Ideas Plus'' award No. IdP2012 000162 to IS.}

\newpage
\centerline{\bf Appendix~A}
\begin{figure}[htb]
\begin{center}
\includegraphics[width=12.5cm,bb=65 65 525 730]{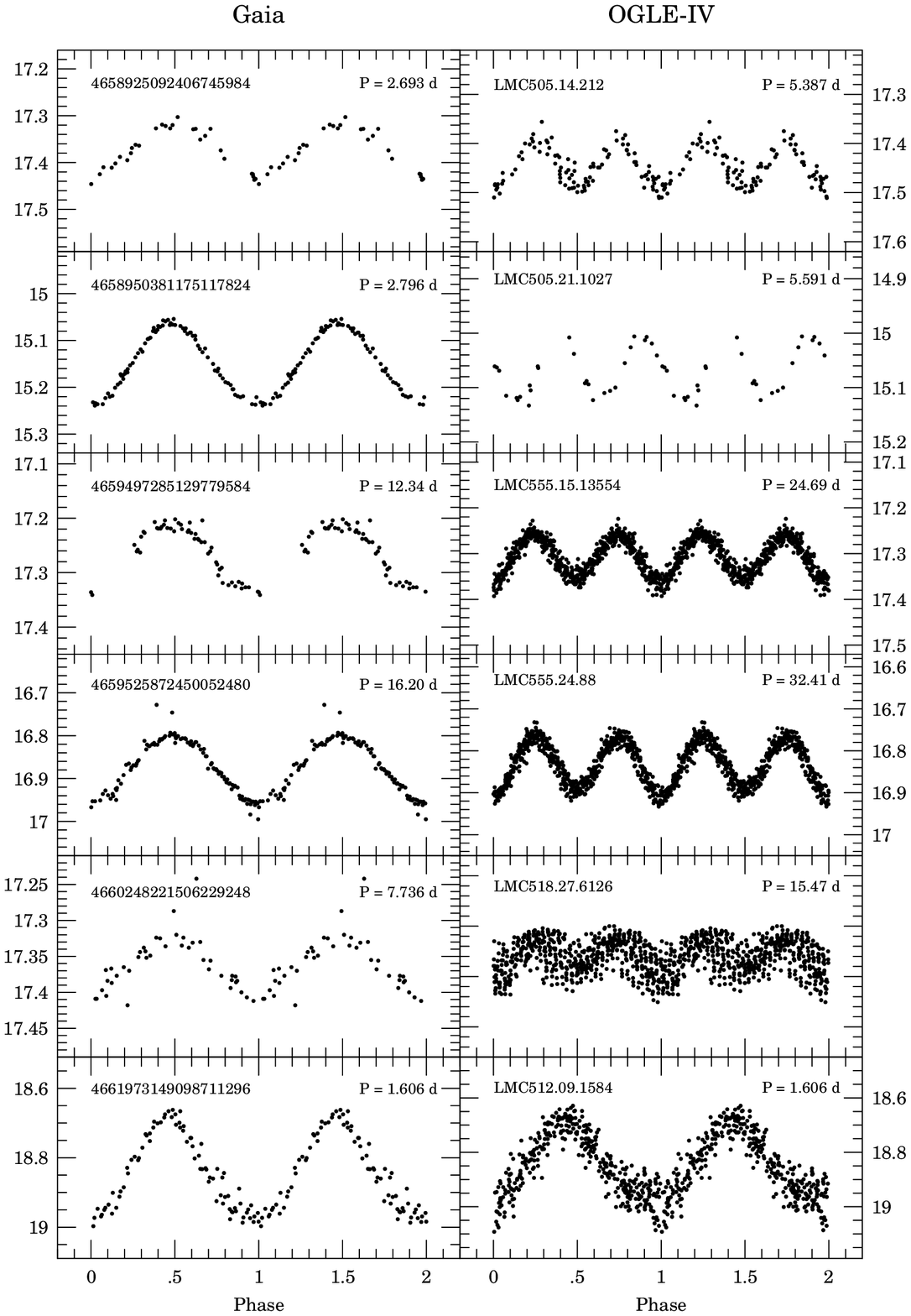}
\end{center}
\FigCap{Gaia and OGLE-IV light curves of misclassified Gaia Cepheids.}
\end{figure}
\begin{figure}[p]
\begin{center}
\includegraphics[width=12.5cm,bb=65 65 525 730]{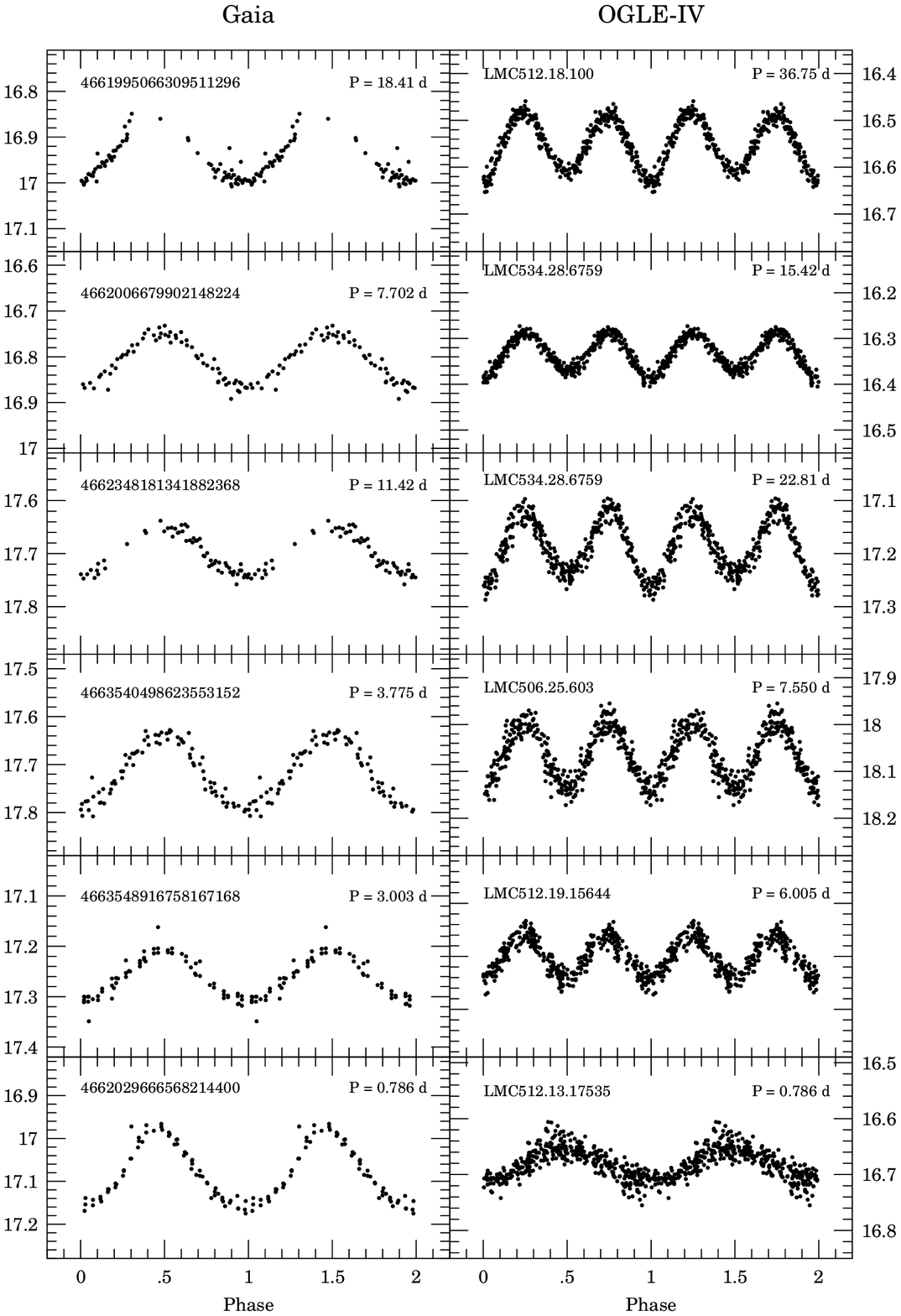}
\end{center}
\FigCap{Gaia and OGLE-IV light curves of misclassified Gaia Cepheids.}
\end{figure}
\begin{figure}[t]
\vglue-5cm
\centerline{\includegraphics[width=12.5cm,bb=70 70 525 580]{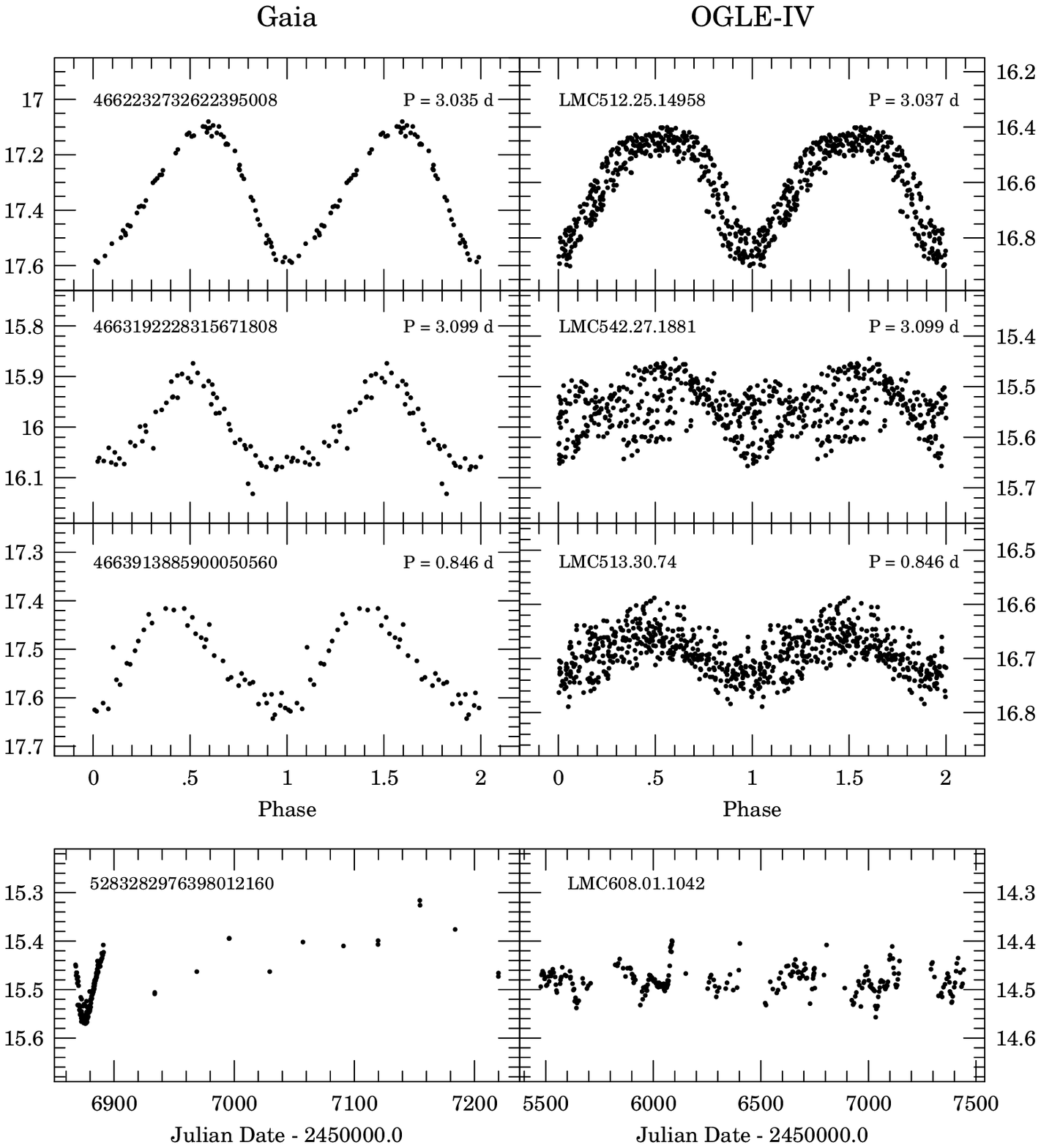}}
\FigCap{Gaia and OGLE-IV light curves of misclassified Gaia Cepheids.}
\end{figure}

\newpage
\begin{figure}[p]
\centerline{\bf Appendix~B}
\vglue0.3cm
\centerline{\includegraphics[width=12.5cm,bb=65 65 525 730]{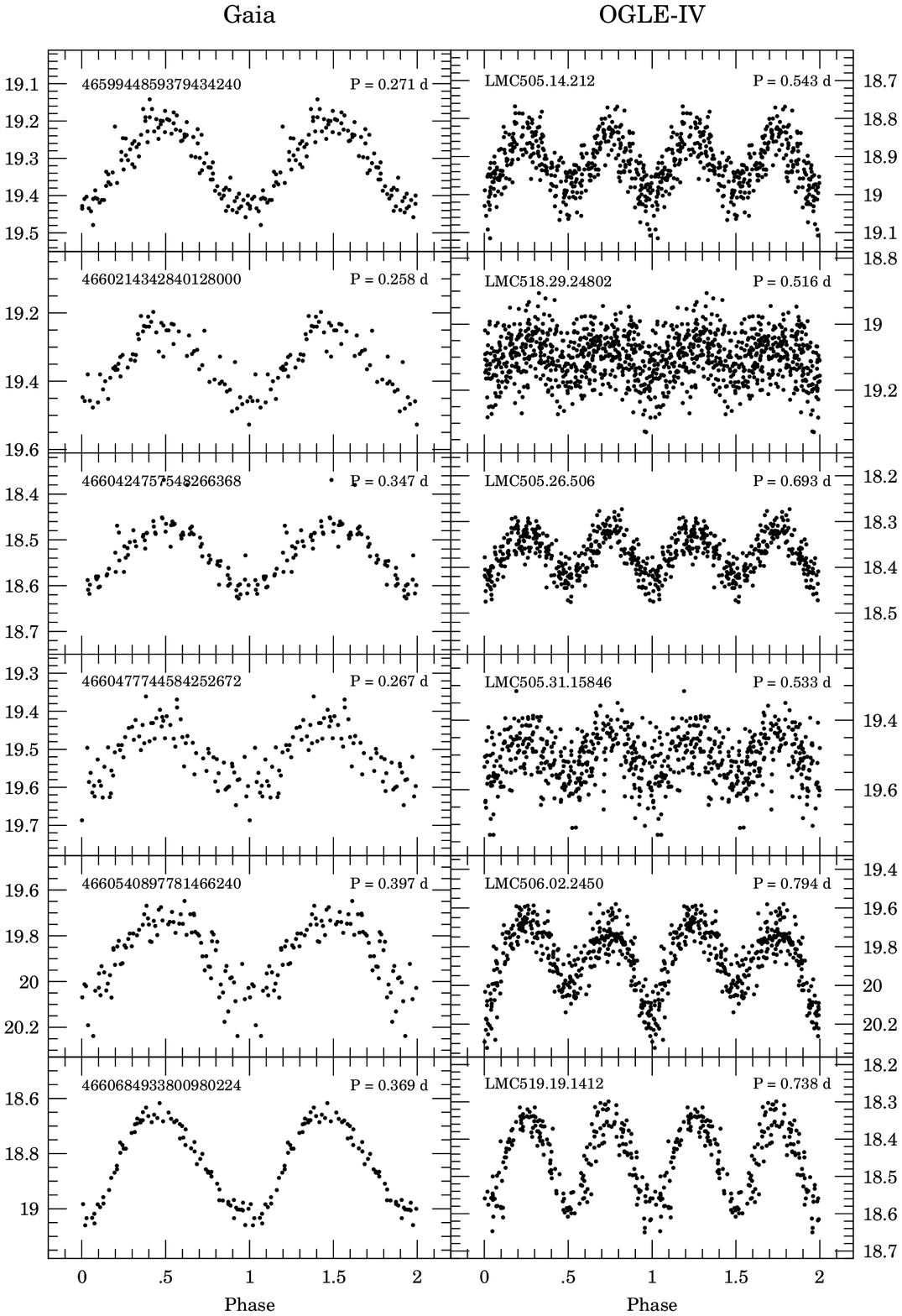}}
\FigCap{Gaia and OGLE-IV light curves of misclassified Gaia RR Lyrae
stars.}
\end{figure}
\begin{figure}[htb]
\begin{center}
\includegraphics[width=12.5cm,bb=65 65 525 730]{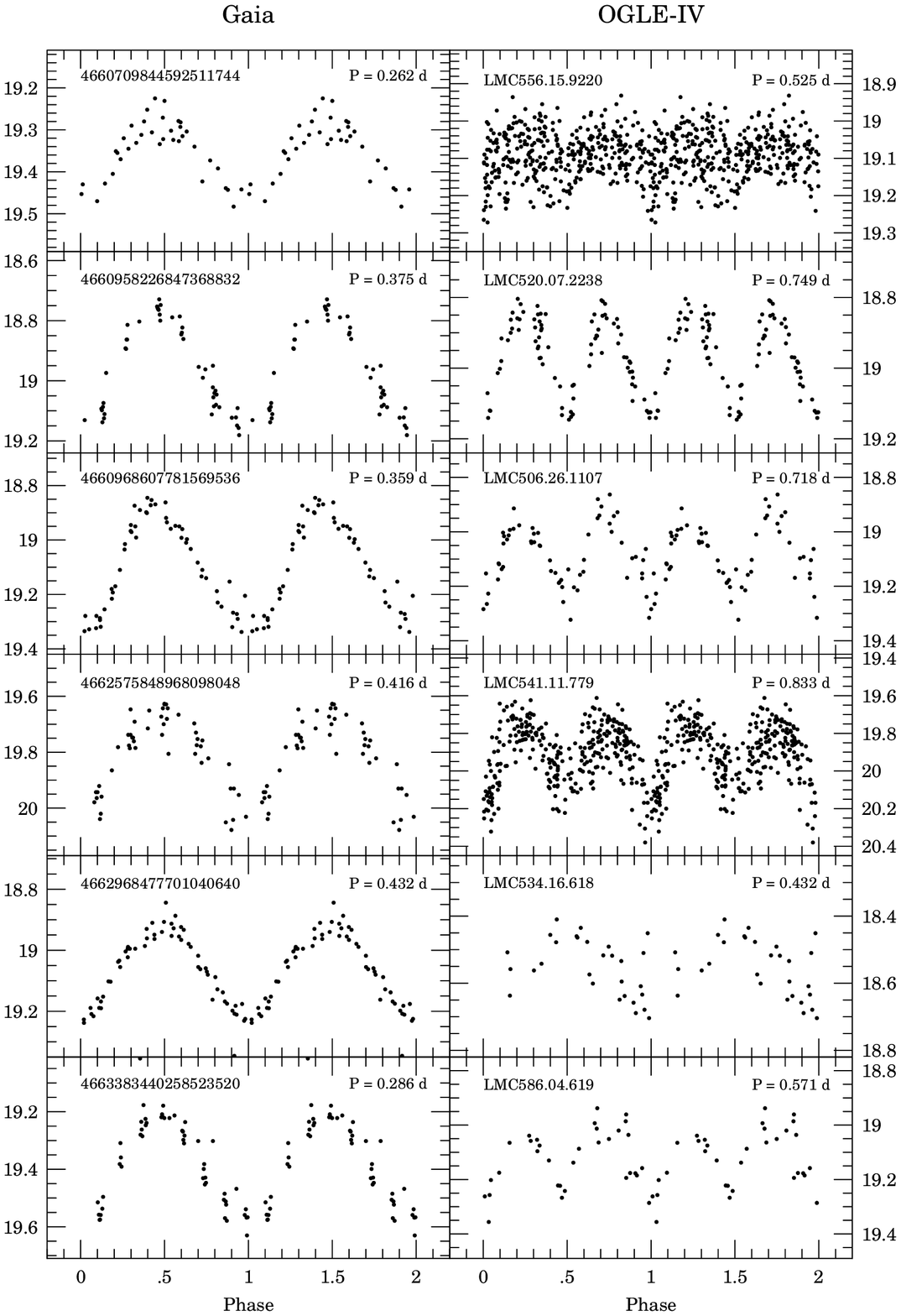}
\end{center}
\FigCap{Gaia and OGLE-IV light curves of misclassified Gaia RR Lyrae
stars.}
\end{figure}
\begin{figure}[htb]
\begin{center}
\includegraphics[width=12.5cm,bb=65 65 525 730]{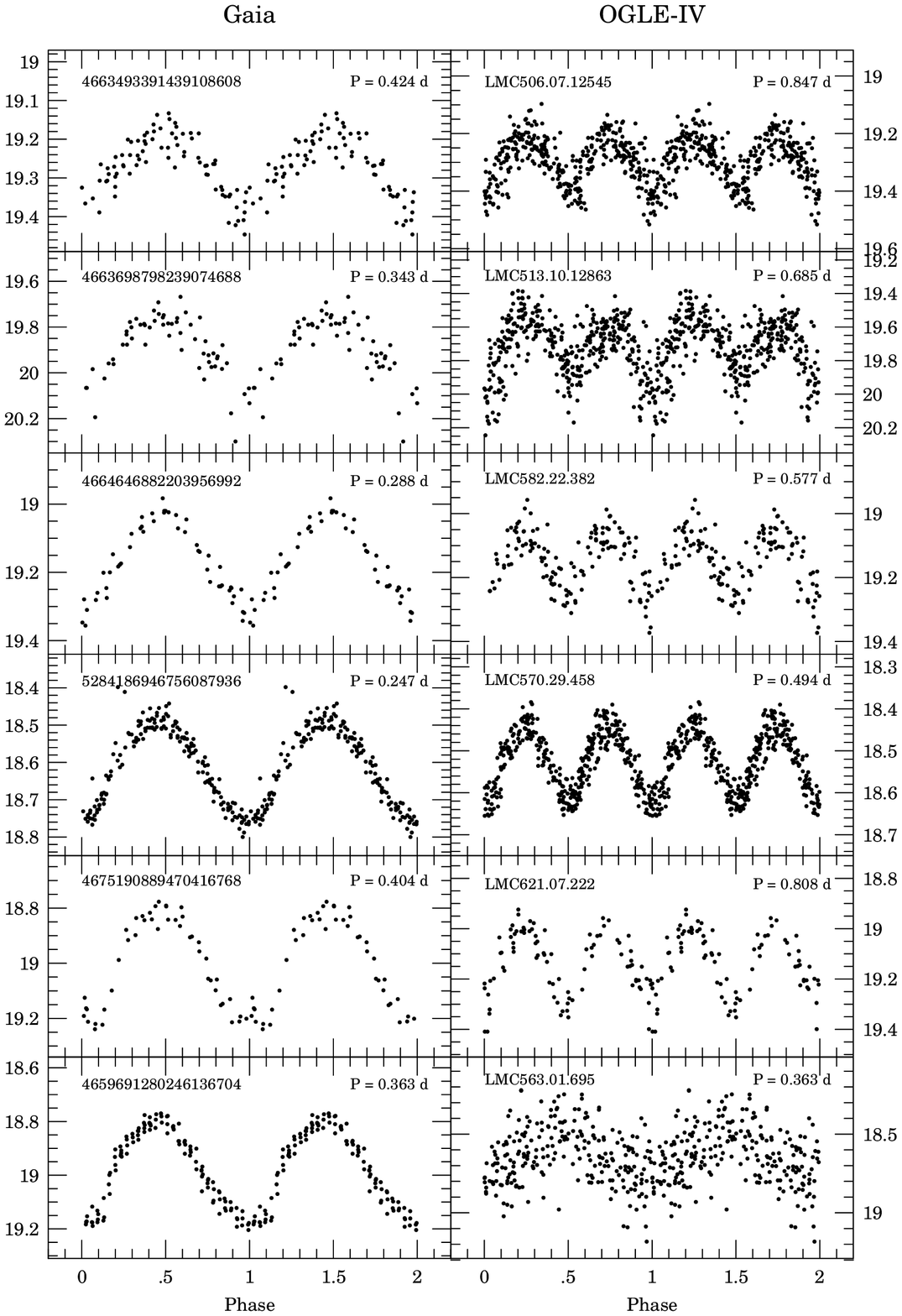}
\end{center}
\FigCap{Gaia and OGLE-IV light curves of misclassified Gaia RR Lyrae
stars.}
\end{figure}

\newpage
\begin{figure}[t]
\vglue-6cm
\centerline{\includegraphics[width=12.5cm,bb=70 60 525 330]{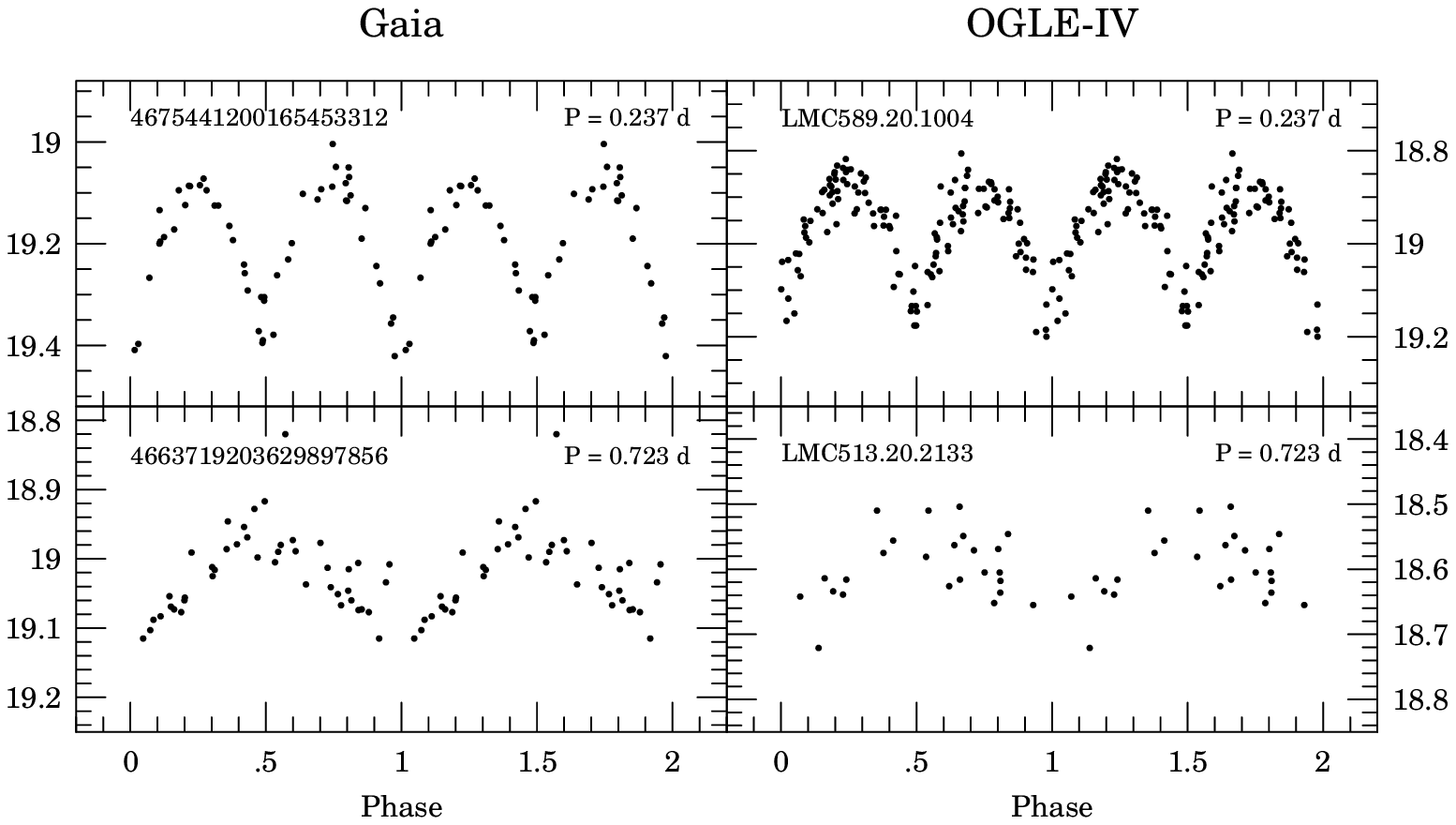}}
\FigCap{Gaia and OGLE-IV light curves of misclassified Gaia RR Lyrae stars.}
\end{figure}


\begin{references}

\refitem{Eyer, L., \etal}{2012}{Ap\&SS}{341}{207}
\refitem{Gaia Collaboration, Prusti, T., \etal}{2016a}{\AA}{~}{in press; arXiv:1609.04153}
\refitem{Gaia Collaboration, Brown A.G.A., \etal}{2016b}{\AA}{~}{in press; arXiv:1609.04172}
\refitem{Gaia Collaboration, Clementini, G., \etal}{2016c}{\AA}{~}{in press; arXiv:1609.04269}
\refitem{Pietrukowicz, P., \etal}{2015}{\ApJ}{813}{40}
\refitem{Soszyñski, I., \etal}{2012}{\Acta}{62}{219}
\refitem{Soszyñski, I., \etal}{2013}{\Acta}{63}{21}
\refitem{Soszyñski, I., \etal}{2014}{\Acta}{64}{177}
\refitem{Soszyñski, I., \etal}{2015a}{\Acta}{65}{233}
\refitem{Soszyñski, I., \etal}{2015b}{\Acta}{65}{297}
\refitem{Soszyñski, I., \etal}{2016}{\Acta}{66}{131}
\refitem{Udalski, A., Szymañski, M.K., and Szymañski, G.}{2015}{\Acta}{65}{1}
\end{references}
\end{document}